\documentclass[12pt]{iopart}
\usepackage{graphicx,epsfig}
\usepackage{amssymb}
\usepackage{iopams}
\usepackage{float}

\def\beq{\begin{equation}}
\def\eeq{\end{equation}} 
\def\br{\begin{eqnarray}}
\def\er{\end{eqnarray}}
\def\benu{\begin{enumerate}}
\def\eenu{\end{enumerate}}

\begin{document}
\def\plotone#1{\centering \leavevmode
\epsfxsize= 0.4\columnwidth \epsfbox{#1}}
\def\plottwo#1#2{\centering \leavevmode
\epsfxsize=.43\columnwidth \epsfbox{#1} \hfil
\epsfxsize=.43\columnwidth \epsfbox{#2}}
\def\plotfiddle#1#2#3#4#5#6#7{\centering \leavevmode
\vbox to#2{\rule{0pt}{#2}}
\includegraphics{#1}}
\newdimen\hhsize\hhsize=.5\hsize
\font\lloyd=cmr8 

\title{Ringing Non-Gaussianity from inflation with a step in the second derivative
of the potential}
\author{Rakhi. R$^{a}$ and Minu Joy$^{b}$ }
\address{$^a$ 
Dept. of Physics, NSS College, Pandalam 689501, India}
\address{$^b$ 
Dept. of Physics, Alphonsa College, Pala 686574, India}
\begin{abstract}
Inflationary model driven by a scalar field whose potential has a step in the second derivative with respect to the field is considered. 
For the best fit potential parameter values, the 3-point function and  the non-Gaussianity associated with the featured  model is calculated. 
We study the shape and scale dependence of the 3-point function.
The distinctive feature of this model is its characteristic ringing behaviour of $f_{NL}$. We can see that the oscillations in $f_{NL}$ 
in this model last for a much longer range of k values, as compared to the previously studied models. In that sense, this model is potentially 
distinguishable from  models with other features in the potential.
\end{abstract}

\maketitle
\section{Introduction}
In standard slow-roll inflation the deviation from Gaussian distribution of the primordial
perturbations \cite{star,mukhanov,lyth4} is predicted to be small. It is of the order of the slow-roll parameters
\cite{mald,seery,chen1,gio}. This result does not hold if the the inflaton undergoes a period of slow-roll
violation during its evolution \cite{chen2,chen3,misao,sriram1}, as can happen if the inflaton potential has
some localized feature \cite{Hu}. The resulting non-Gaussianity \cite{2020} then becomes shape and scale
dependent and modes that exit Hubble scale around the time the field cross the feature
can pick up large non-Gaussianities.
An inflationary model where the inflaton potential has a feature in its
second derivative with respect to the inflaton has been proposed in \cite{minu}. In that paper, the
authors showed that the power spectrum picks of small oscillations superimposed on a
flat spectrum and the spectral index can experience a jump around the feature.
This scenario helps to explain the local running in the spectral index observed in the
WMAP and Planck data \cite{wmap,Planck2015Inf,Planck2018Inf}.  
In the present work, we are interested in studying the non-Gaussianity predicted by this
model. The analysis of the running of non-Gaussianity using Planck data is 
given in \cite{Planck2018NG}, in the context of some well defined inflationary models. Inflationary models that 
predicts a mildly scale dependent bispectrum, termed as the running of the bispectrum \cite{chen4,Byrnes,shand} is discussed there.   
In the present work, we compute the 3-point function of the curvature perturbation and study the shape and scale dependence of the 3-point function.
The paper is organized as follows: in section 2 we describe our model and give the
background evolution and the resulting power spectrum of curvature perturbations. In
section 3 we compute the three-point function and the non-Gaussianity from the model and then conclude by summarizing our results in section 4.

\section{The model with a step in the second derivative of the potential}
In \cite{minu} we showed that the well-known model with a Higgs-like 
potential which is used in the hybrid inflationary scenario too,

\beq V(\psi,\phi) = \frac{1}{4\lambda}\left (M^2 -
\lambda\psi^2\right )^2 + \frac{1}{2}m^2\phi^2 +
\frac{g^2}{2}\phi^2\psi^2~, \label{eq:hybrid} \eeq
could successfully give rise to a step in the primordial spectral index.  
At $\phi_c = M/g$ the curvature of $V(\psi,\phi)$ along the $\psi$ direction
vanishes so that $m_\psi^2  > 0$ for $\phi > \phi_c$ while $m_\psi^2  < 0$ 
for $\phi < \phi_c$. This implies that for large values of the inflaton 
$\phi$ the auxiliary field $\psi$ rolls towards $\psi = 0$.
However, once the value of $\phi$ falls below $\phi_c$ the $\psi = 0$ 
configuration is destabilized resulting in a rapid cascade (mini-waterfall) 
which takes $\psi$ from $\psi = 0$ to its minimum value.
This potential has four parameters, namely, M, m, g and $\lambda$.  The model confrontation with WMAP–7 data 
has been done in \cite{minu2} and the best fit values of potential parameters were obtained with N = 40 and N = 60, 
where N represents the number of e-folds after the phase transition. Table below gives the potential paramters best fit with
the Bicep-Keck-Planck likelihoods (combined BICEP2 and Keck Array October 2018 data in combination with the 2018 Planck data)\cite{cosmomc,comb1,comb2}.


\begin{table}[ht]
\begin{center}
\begin{tabular}{|l|l|l|}
 \hline
Parameter & Values for N=60 & Values for N=40\\\hline
$M/m_{pl}$ & 7.43$\times$ $10^{-4}$ & 8.22 $\times$ $10^{-4}$\\ \hline
$m/m_{pl}$ & 4.6$\times$ $10^{-7}$& 6.9$\times$ $10^{-7}$ \\ \hline
$g$ & 2.77 $\times$ $10^{-4}$ & 3.75 $\times$ $10^{-4}$ \\ \hline
$\lambda$ & 0.1 & 0.1 \\\hline
\end{tabular}
\end{center}
\caption{\footnotesize {Best fit values of Potential Parameters}}
\end{table}


\subsection{Background Evolution}
With this potential, by initiating evolution with initial field value $\phi_i > \phi_c$, the coupled
system of equations of the background inflaton and scale factor of expansion of spacetime
exhibits inflation with the inflaton rolling the potential, till inflation ends and $\phi$
finally oscillating about the potential minimum.
The slow-roll conditions $\epsilon$, $\eta < 1$, where
$\epsilon$, $\eta$ are given by
\[
 \epsilon\equiv 3 \frac{\dot{\phi}^2/2}{\dot{\phi}^2/2+V}, \hspace{2cm} \eta\equiv-3\frac{\ddot{\phi}}{3H\dot{\phi}}
\]
are satisfied throughout the inflationary period. However, $\eta$ has a discontinuity at $\phi= \phi_ c$ since
it is proportional to $V''$. 
We explore the consequences of this
discontinuity in the behaviour of the two-point and three-point functions of
the perturbations of the dynamical variables. 

\subsection{Two-point function and power spectrum of scalar perturbations}
The Fourier modes of the curvature perturbation satisfies the equation
\begin{equation}
 {\cal R}^{\prime\prime}_k + 2 \frac{z^\prime}{z}{\cal R}^\prime_k + k^2 {\cal R}_k =0
\end{equation}
where the prime denotes derivative with respect to conformal time and the quantity $z$ is given by
\begin{equation}
 z\equiv \frac{a}{H}\sqrt{\rho + p}=\frac{a\dot\phi}{H}
\end{equation}
${\cal R}$ is related to the Mukhanov-Sasaki variable u as ${\cal R} = u/z$. The scalar power spectrum is then defined as
\begin{equation}
 {\cal P}_s(k)=\frac{k^3}{2\pi^2}|{\cal R}_k |^2
\end{equation}
with the amplitude of the curvature perturbation ${\cal R}_k$ evaluated, in general, at the end of
inflation.

\begin{figure}[!h]
\centerline{
\psfig{figure=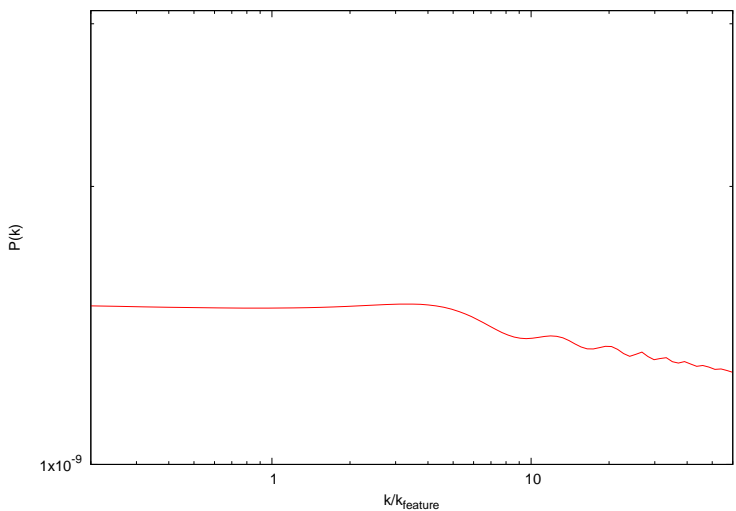,width=0.45\textwidth,angle= -90} }
\caption{\small Primordial spectrum for a model with step in the second derivative of the potential
}
\label{fig:pk60}
\end{figure}

Under slow-roll, the power spectrum is approximately given by
\begin{equation}
 {\cal P}_s(k)\simeq\frac{1}{2\epsilon}\frac{H^2}{2\pi}\left(\frac{k}{aH}\right)^2
\end{equation}
with $n_s$ given in terms of slow-roll parameters as
\begin{equation}
 n_s \simeq 1-2\epsilon -\eta
\end{equation}
For the potential given by Eq.(\ref{eq:hybrid}), one can immediately notice that $n_s$ will have discontinuity due to its dependence on $\eta$. 
This will result in the power spectrum having a jump in its slope at a scale set by $\phi_c$. The full analytic expression for the
power spectrum has small oscillations superimposed around the scale of change of slope, as shown in \cite{minu}. Figure \ref{fig:pk60} shows the quasi
flat $\mathcal{P}_k$ for this mini waterfall hybrid model.

\section{Three-point function and non-Gaussianity from the model}
Our approach is based on the numerical evaluation of both the perturbation equations and the integrals which
contribute to the 3-point function as described by Chen \textit{etal} in \cite{chen2, chen3}.
To compute the 3-point correlation function, one can substitute the mode solutions
$u_k$ into equation (3.17){\footnote{$
H_{int}(\tau) =
-\int d^3x \Bigg\{ a \epsilon^2 \zeta \zeta'^2  + 
a\epsilon^2 \zeta (\partial \zeta)^2  - 
2 \epsilon \zeta' (\partial \zeta) (\partial \chi)
+ \frac{a}{2} \epsilon \eta' \zeta^2 \zeta'
+ \frac{\epsilon}{2a} (\partial \zeta) (\partial \chi) 
(\partial^2 \chi)
+ \frac{\epsilon}{4a} (\partial^2 \zeta) (\partial \chi)^2
\Bigg\} $
}} of \cite{chen2}, and integrate the mode functions from $\tau_0$\footnote[1]{{$\tau_0$ is an 
arbitrary time when all three modes are well inside the horizon.}}
through to the end of inflation. This integral can be done semi-analytically for simple models, provided the
slow-roll parameters are small and relatively constant.  For standard single field slow-roll inflation, the terms of
order $\epsilon^2$ in the above mentioned equation are the dominant contributors to the
3-point function and the other terms of order $\epsilon\eta'$ and
$\epsilon^3$ were neglected in
Refs.~\cite{mald,seery,chen1}. 
In the presence of a step in the potential, the $\epsilon
\eta'$ term becomes large and dominant \cite{chen2,chen3}. The step in the second order derivative of
the potential will make the $\epsilon\eta'$ term much higher compared to
$\epsilon^2$ term and hence leads to a modification of the standard slow-roll results. 
Thus the term of our particular interest is the $\epsilon\eta'$ term

 \[
  I_{\epsilon \eta^\prime} \propto i \left(\prod_i u_i(\tau_{end})\right)
\int_{\infty}^{\tau_{end}}d\tau a^2\epsilon\eta^\prime 
\left(u_1^*(\tau)u_2^*(\tau)\frac{d}{d\tau}u_3^*(\tau)
+ \hspace{0.2cm} 
two ~perm\right) \times
\]
\begin{equation}
 \hspace{3cm}\left(2\pi\right)^3  \delta^3
\left(\sum_i \textbf{k}_i\right) + c.c
\label{eq5.13}
\end{equation}
where the “two perm” stands for two other terms that are symmetric under permutations of the indices 1, 2 and 3, where 1, 2, 3 are shorthand for $k_1$, $k_2$ and $k_3$.
For our step model, $\epsilon\eta'$ can be written as, 

\begin{equation}
\epsilon\eta' = 6aH\left( 2\epsilon^2 - \frac{\epsilon\eta}{2} +
\frac{5}{6}{\epsilon^2\eta}
-\frac{2}{3}\epsilon^3 - \frac{\epsilon\eta^2}{12} - \epsilon\frac{V_{\phi\phi}}{3H^2}\right)
\end{equation}


In order to integrate Eq. (\ref{eq5.13}) numerically, we follow the procedure detailed in \cite{chen3}. 
For the present model with a step in the second deriavative of the potential, 
\begin{equation}
 \nu_k(\tau_0) =
\frac{\sqrt{\pi\tau}}{2}H_{\mu_1}^{(2)}(k\tau_0)~, 
\label{eq5.17}
\end{equation}
where $H_{\mu_1}^{(2)}(k\tau_0)$ is the Hankel function and ${\mu_1} =
\frac{3}{2} - \frac{V''_-}{3H_0^2} + 3 \epsilon_0 $, where $V''_-
\equiv \left(\frac{d^2V}{d\varphi^2}\right)_{\rm before\hspace{0.1cm} phase \hspace{0.1cm}transition}$. 

Every 3-point correlation function has two main attributes: shape and scale. 
Following \cite{chen2,chen3} we define the parameter,
$\cal G$ to describe non-Gaussianities with both shape and scale dependence:
\begin{equation}
 \frac{{\cal G} (k_1, k_2, k_3)}{k_1 k_2 k_3}\equiv \frac{1}{\delta^3 (\textbf{k}_1 + \textbf{k}_2 + \textbf{k}_3)}
\frac{(k_1 k_2 k_3)^2}{P_k^2(2\pi)^7}\langle\zeta(\textbf{k}_1)\zeta(\textbf{k}_2)\zeta(\textbf{k}_3)\rangle
\label{eq5.14}
\end{equation}

In the absence of the sharp feature, Eq. (\ref{eq5.14}) reduces to the local form with 
\begin{equation}
 {\cal G} = (3/10)f_{NL}^{local} \sum k_i^3
\label{eq5.15}
\end{equation}
Using Eqs. (\ref{eq5.14}) and (\ref{eq5.15}), 
we can calculate the non-Gaussianity parameter, $f_{\rm NL}^{\rm local}$, for our model.
 
\begin{equation}
\langle \zeta(\textbf{k}_1)\zeta(\textbf{k}_2)\zeta(\textbf{k}_3)
 = (2 \pi)^7\delta^3(\textbf{k}_1 + \textbf{k}_2 + \textbf{k}_3)
\left(-\frac{3}{10}f_{NL} \mathcal{P}_k^2  \right) \frac{\Sigma_i k_i^3}{\Pi_i k_i^3} 
\end{equation}

\begin{figure}[!ht]
\centerline{
\psfig{figure=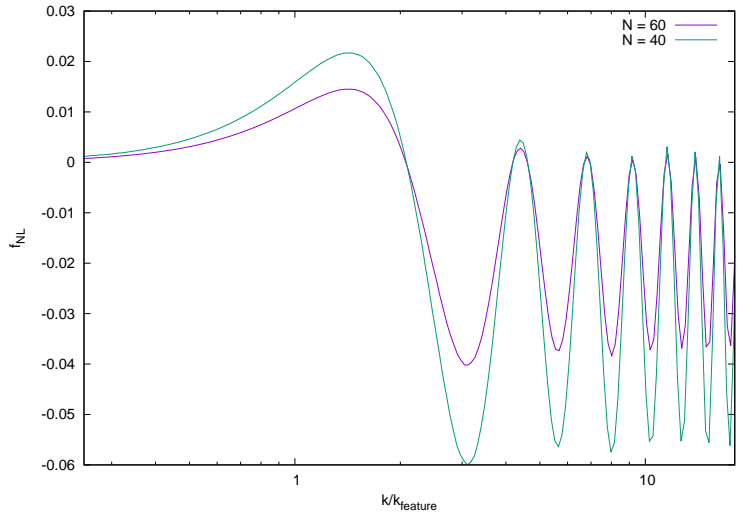,width=0.45\textwidth,angle= -90} }
\caption{\small
$f_{NL}^{equil}$ for the equilateral case. x axis is $k/k_{feature}$ 
}
\label{eq}
\end{figure}

Since this model has already been used to improve the fit between LCDM cosmology and the observed power spectrum, we can take the best fit
potential parameter values and compute the resulting 3-point function. Figure \ref{eq} gives the $f_{NL}^{equil} $ for our model, for
 the equilateral configuration ($k_1=k_2=k_3$). Purple line is for N = 60 and green for N = 40. 
 The numerical value is not much larger than those in standard single filed, slow-roll inflation. The distinctive feature of this 
non-Gaussianity is its characteristic ringing behaviour of $f_{NL}$. 

We can see that the oscillations in $f_{NL}$ in this model last for
a much longer range of k values, as compared to the previously studied models \cite{chen2,chen3}. In that sense, this model is 
potentially distinguishable from  models with other features in the potential. 
 

\section{Conclusion}
The single filed, slow-roll models of inflation generically yields a negligible primordial non-Gaussianity,
which is not even observable. Thus the bispectrum analysis of CMB data can be considered as a promising candidate 
to discriminate between the degenerate inflationary models. In the present work, we considered a variant of hybrid
inflation where the potential has a discontinuity in its second derivative with respect to field. This describes a 
fast second order phase transition during inflation that occurs in some other scalar field weakly coupled to the inflaton. 
The 3-point correlation function is numerically integrated for this anomalous inflationary model where slow-roll is 
violated for a brief moment. The transient violation of the slow-roll leads to an oscillating and scale dependent 3-point
function.  For the typical potential parameter values, non-Gaussianity associated with the featured potential model  
is  found to be oscillating.  

The distinctive feature of this non-Gaussinaity is its characteristic ringing behaviour; $f_{NL}$ oscillates between a maximum and minimum value.
The oscillations in $f_{NL}$ in this model last for
a much longer range of k values, as compared to the previously studied models. In that sense, this model is 
potentially distinguishable from  models with other features in the potential.  


\ack
We acknowledge the use of high performance computing system at  IUCAA. We thank Pravabati Chingangbam for contributions during the early stages
 of this project. MJ acknowledges the Associateship of IUCAA.

\end{document}